\documentclass{article}
\usepackage{arxiv}
\usepackage{cite}
\usepackage{amsmath,amssymb,amsfonts}
\usepackage{algorithmic}
\usepackage{graphicx}
\usepackage{textcomp}
\usepackage{breqn}
\usepackage[]{algorithm2e}
\usepackage{caption}
\usepackage{subcaption}
\usepackage[utf8]{inputenc}
\usepackage{lipsum}
\usepackage{scalefnt}
\usepackage{stfloats}

\def\BibTeX{{\rm B\kern-.05em{\sc i\kern-.025em b}\kern-.08em
    T\kern-.1667em\lower.7ex\hbox{E}\kern-.125emX}}
\begin{document}
\title{Cross-Database and Cross-Channel ECG Arrhythmia Heartbeat Classification Based on Unsupervised Domain Adaptation}
\author{Md Niaz Imtiaz,  and Naimul Khan}
\thanks{The manuscript was submitted on May 24, 2023. This work was supported by the Natural Sciences and Engineering Research Council of Canada (NSERC) and the Government of Canada's New Frontiers in Research Fund (NFRF).}
\thanks{Md Niaz Imtiaz and Naimul Khan are with the Department of Electrical, Computer and Biomedical Engineering, Toronto Metropolitan University, Toronto, ON M5B 2K3, Canada (e-mail: niaz.imtiaz@torontomu.ca; n77khan@torontomu.ca).}

\maketitle

\begin{abstract}
The classification of electrocardiogram (ECG) plays a crucial role  in the development of an automatic cardiovascular diagnostic system. However, considerable variances in ECG signals between individuals is a significant challenge. Changes in data distribution limit cross-domain utilization of a model. In this study, we propose a solution to classify ECG in an unlabeled dataset by leveraging  knowledge obtained from labeled source domain. We present a domain-adaptive deep network based on cross-domain feature discrepancy optimization. Our method comprises three stages: \textit{pre-training}, \textit{cluster-centroid computing}, and \textit{adaptation}.  In pre-training, we employ a Distributionally Robust Optimization (DRO) technique to deal with the vanishing worst-case training loss. To enhance the richness of the features, we concatenate three temporal features with the deep learning features. The cluster computing stage involves computing centroids of distinctly separable clusters for the source using true labels, and for the target using confident predictions. We propose a novel technique to select confident predictions in the target domain. In the adaptation stage, we minimize compacting loss within the same cluster, separating loss across different clusters, inter-domain cluster discrepancy loss, and running combined loss to produce a  domain-robust model. Experiments conducted in both cross-domain and cross-channel paradigms show the efficacy of the proposed method. Our method achieves superior performance compared to other state-of-the-art approaches in detecting ventricular ectopic beats (V), supraventricular ectopic beats (S), and fusion beats (F). Our method achieves an average improvement of 11.78\%  in overall accuracy over the non-domain-adaptive baseline method on the three test datasets.
\end{abstract}

\section{Introduction}
\label{sec:introduction}
According to research by the UN \cite{mendis2011global}, cardiovascular illness is now the leading cause of mortality worldwide. Studies indicate that 80\% of sudden cardiac fatalities have a strong correlation with arrhythmia \cite{mehra2007global}. Therefore, a timely diagnosis of arrhythmia is needed, and it must be done accurately. The identification of cardiac arrhythmias heavily relies on the electrocardiogram (ECG), a physiological signal that provides information about the electrical activity of the heart. Furthermore, the prevention, diagnosis, and treatment of cardiovascular illnesses all depend on high-precision automatic diagnostics. The classification of heartbeat is a fundamental and common task in the automated detection of arrhythmias.

Over the past few years, several machine learning and deep learning approaches have been put forth to identify various heart irregularities using ECG signals. Conventional methods employ several classification algorithms, like  Support Vector Machines \cite{yang2018automatic}, and K-Nearest Neighbor \cite{raj2018sparse}, on hand-crafted features. Several techniques for classifying ECG heartbeats have been developed as a result of recent advancements in deep learning. These methods leverage the potent feature-learning capabilities of deep learning algorithms and large volumes of annotated clinical data. The proposed deep learning methods can be broadly categorized into two types: convolutional neural networks (CNN) \cite{li2019automated,zhai2018automated,kiranyaz2015real,xu2020ecg,sellami2019robust} and recurrent neural networks (RNN) \cite{zhang2017patient, singh2018classification, rana2019ecg}. CNNs employ several convolutions with various filters to directly extract key features from the ECG data. These high-level features are then forwarded to classifiers for prediction. On the other hand, RNNs evaluate the temporal relationships within ECG data and incorporate features at various time steps. 

Even though deep learning techniques have advanced significantly, they still face challenges in the cross-domain paradigms. Individual differences have a significant impact on the morphological properties of ECG signals. For this reason, when evaluated on new patient data, models perform significantly worse. Domain shift \cite{quionero2009dataset}, which refers to differences between test and training data,  may go against the fundamental identically distributed assumption in a learning-based scheme. Moreover, deep learning models require an extensive amount of labeled data for training to achieve good results. In the real world, collecting sufficient amounts of labeled ECG data is typically costly and laborious. The ECG signal takes a very long time to capture, and the changes are subtle \cite{goldberger2017clinical} \cite{acharya2007advances}, making manual labeling exceedingly time-consuming. Also, in some real-world settings, it is hard to obtain the data labels of ECG signals collected from new participants, which prevents us from re-training new supervised models for these cases. Therefore, it is a challenging but necessary endeavor to achieve precise cross-domain ECG heartbeat classification. 

Our solution to the problems discussed above is a domain-adaptive deep learning model based on cluster discrepancy optimization to classify arrhythmia heartbeats without additional annotation from experts. The model comprises a feature extractor based on residual blocks. Inspired by Ye et al.'s study \cite{ye2022ecg}, we add a bi-classifier after the feature extractor. A bi-classifier minimizes inconsistencies in the predictions made by a single classifier. In a bi-classifier network, the outputs of the two classifiers are combined to get the final prediction. The drawback of a bi-classifier network is that combining the two outputs could result in an incorrect prediction, even if one of the classifiers has predicted the correct label. So, proper training to minimize the discrepancy between classifiers is required. The proposed method is composed of three stages: pre-training, computing cluster centroids, and adaptation. The pre-training stage trains the model using source data and labels to correctly classify the ECG segments by minimizing the classification loss and the discrepancy loss between the two classifiers. Taking inspiration from Sagawa et al.'s study \cite{sagawa2019distributionally}, we incorporate a distributionally robust optimization (DRO) method during the training process. The DRO method tackles the issue of disappearing worst-case training loss. Although reducing the vanishing worst-case training loss enhances the accuracy of predictions, the DRO method is susceptible to issues when there is a large discrepancy between source and target distributions. During the computing cluster centroids and adaptation stage of our method, our objective is to minimize the distribution differences between the source and target distributions. The cluster centroids of the source and target domains are computed in the second stage. These centroids are used to compact samples within the same cluster and to move the clusters farther apart from each other. Finally, in the adaptation stage, the feature distribution differences between the source and target domain are minimized through four loss functions. 

 Our approach is capable of enhancing the performance of deep learning models on new data without the need for any supplementary human effort. Unlike domain-specific techniques that maintain a distinct model for each domain, our approach employs a single global model for all test domains. The suggested approach is appropriate for applications that demand efficient adaptation to new data from diverse distributions, such as customized portable devices and  online diagnostic systems. 

 This paper's key contributions are as follows:

  (1) A novel technique to select confident predictions in the unlabeled target domain is proposed, which in turn improves the precision of cluster separation.

  (2) To mitigate the discrepancy in feature distributions among domains, two new objective functions are introduced: the \textit{running combined loss} and the \textit{inter-domain cluster discrepancy loss}. These objective functions are utilized alongside two existing ones \cite{wang2021inter}, namely the \textit{cluster-compacting loss} and the \textit{cluster-separating loss}, during the adaptation stage.
  
  (3) Two-stage training after pre-training is performed to efficiently organize distinguishable clusters in the source domain first and then use them to minimize the cluster discrepancy between the source and target domains.
 
Inspired by Niu et al. \cite{niu2020deep}, we also combine three time features with the deep features to improve the performance of our proposed approach further. The efficacy of our method is demonstrated through experimental results conducted on public databases. The MIT-BIH Arrhythmia Database (MITDB) \cite{goldberger2000physiobank} is used to train the proposed model, while the St. Petersburg INCART 12-lead Arrhythmia Database (INCARTDB) and the European ST-T Dataset (ESTDB) \cite{goldberger2000physiobank} are used to test it. This study considers both cross-database and cross-channel paradigms. Our proposed method is compared against five recent approaches \cite{sun2016deep,sagawa2019distributionally,huang2020self,wang2021inter,niu2020deep} that are recognized for their high performance. This comparison is made by evaluating them using the same network architecture and experimental setting as our proposed method. Our proposed technique achieves the overall accuracy of 84.61\%, 82.32\%, and 76.44\% on INCARTDB (cross-domain paradigm), INCARTDB (cross-domain and cross-channel paradigm), and ESTDB, respectively, which is considerably higher than other approaches. An ablation analysis and comparison of our proposed method with the method without domain adaptation are shown in the experimental results section.

\section{Related works}
\label{sec:related works}

Unsupervised domain adaptation intends to recognize the unlabeled target data by transferring the deep feature knowledge obtained from the labeled source data. Sagawa et al. suggested a group distributionally robust optimization algorithm, which necessitates the samples to be explicitly annotated with their respective groups \cite{sagawa2019distributionally}. Their model was trained to minimize the loss that would occur in the worst-case scenario over groups present in the training data. Sun and Saenko extended the idea proposed by Sun et al. \cite{sun2016return} to propose a deep correlation alignment (CORAL) method to handle situations where the target domain is unlabeled\cite{sun2016deep}. By means of a linear transformation, the CORAL technique adjusts the second-order statistics of the source and target distributions to match each other. Their approach involves integrating CORAL directly into deep networks by creating a differentiable loss function that reduces the gap between the correlations of the source and target. A heuristic training technique called representation self-challenging (RSC) was introduced by Huang et al., which considerably enhances the ability of CNNs to generalize to out-of-domain data \cite{huang2020self}. Through an iterative process, RSC eliminates the dominant features that are activated on the training data and compels the network to activate the remaining features that are correlated with labels. Their method seems to provide feature representations that are useful for handling out-of-domain data, without requiring any knowledge about the new domain beforehand or the need to learn additional network parameters.

Several techniques for domain adaptation have been proposed to work with ECG data. Niu et al. suggested an adversarial domain adaptation-based deep learning approach for classifying ECGs \cite{niu2020deep}. In their model, the high-level features obtained by the feature extractor are passed through a domain discriminator module and a classifier module in parallel. The domain discriminator module resolves the issue of insufficient model depth and low-feature abstraction. The classifier module combines the temporal features with the extracted high-level features to increase feature diversity. A multi-source domain generalization model for ECG classification was developed by Hasani et al. to handle the distribution discrepancy issue that arises when data are collected from numerous sources under various acquisition situations \cite{hasani2020classification}. They used a combined convolutional neural network (CNN) and long short-term memory (LSTM) model to obtain features and the adversarial domain generalization technique to avoid the inconsistency between the training and test data. They also used a variety of augmentation techniques, such as lead dropout, random ECG padding and cropping, and introducing low-frequency aberrations, to boost generalization.

 Wang et al. introduced a domain adaptive ECG arrhythmia classification (DAEAC) method to enhance the deep neural network's performance in the inter-patient paradigm\cite{wang2021inter}. To reduce the distribution differences between the data used for training and testing, they introduced two loss functions named cluster-aligning loss and cluster-maintaining loss. A subdomain adaptive deep network (SADN) was introduced by Jin YR et al. where they excavated the detection knowledge from labeled source domain data and used the knowledge to enhance performance on unlabeled target domain data \cite{jin2022multi}. They used convolutional layers, residual blocks,  and squeeze-and-excitation-residual blocks for automatically extracting significant deep features. To limit data distribution disagreement across datasets, they used a loss function that incorporates the concept of local maximum mean discrepancy. 

Although a few attempts were made for ECG arrhythmia classification across different domains with distribution disparities, they still suffer from unsatisfactory performance for different types of arrhythmias. Inspired by the analysis and considerations mentioned above, this paper proposes a novel deep adaptive model that utilizes knowledge gained from labeled source domain data to enhance classification accuracy on unlabeled target domain data. While some recent domain-adaptation-based methods have presented results on ECG classification \cite{niu2020deep,jin2022multi,hasani2020classification}, they either utilize different groups as source-target within the same database or use nonpublic databases. None of them have utilized the same train-test configuration as this study. We evaluate recent approaches against our proposed method by employing identical network architecture and train-test configuration.

\section{Proposed Method}
\label{sec:proposed method}
\subsection{Framework}
The framework of the proposed approach is demonstrated in Fig. 1. In general, we use the term \textit {source domain} to refer to the training dataset and \textit {target domain} to refer to the test dataset. Assume we have labeled source data $X_s$= $\{{x_s^i}\}_{i=1}^{N_s}$ and their corresponding labels $Y_s$= $\{{y_s^i}\}_{i=1}^{N_s}$, as well as unlabeled target data $X_t$= $\{{x_t^i}\}_{i=1}^{N_t}$. Our objective is to learn a function $F$ using both labeled source data and unlabeled target data, so that it can predict the labels of target data with high accuracy. 

\begin{figure*}[!t]
    
    \centerline{\includegraphics[width=0.8\textwidth]{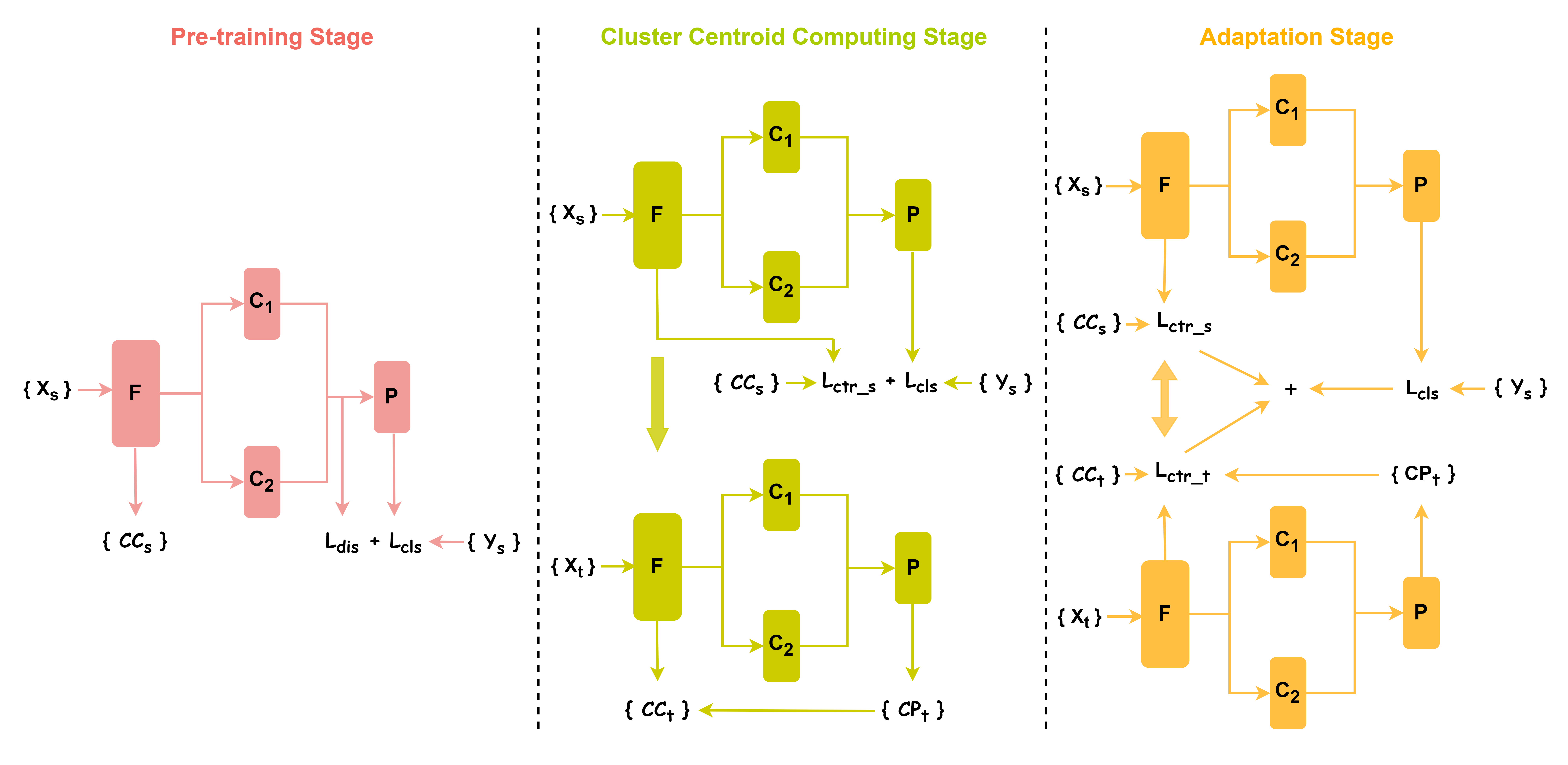}}
    \caption{The framework of the proposed domain-adaptive model.}
    \label{framework}
\end{figure*}

We propose a network composed of a feature extractor and two parallel classifiers (Fig. 2). Feature extractor takes ECG segments and automatically obtains distinctive deep features. The feature extractor is composed of three residual blocks and three max-pooling layers. Each residual block has three 1D convolution layers. To generate deep feature maps, the input layer undergoes the application of the first and second convolution layers. Similarly, the third convolution layer is employed on the input layer to produce shallow feature maps. The residual blocks are responsible for compressing the input vector's length and obtaining deep  features for the classification process that follows. The deep features are then passed through two parallel classifiers. In situations where a single classifier produces inaccurate predictions despite the feature extractor generating good distinctive features, a bi-classifier can rectify the problem. Moreover, we use the discrepancy between the two classifiers to identify confident predictions in the target domain. Each classifier has three fully connected layers. Before the last fully connected layer of the classifier, three corresponding time features (described in the \textit {data preprocessing and network inputs} section) are added to the deep features, which enhances the feature diversity. The outputs of the two classifiers are then combined to get the predicted heartbeat category. 

\begin{figure*}[!t]
    \centerline{\includegraphics[width=0.8\textwidth]{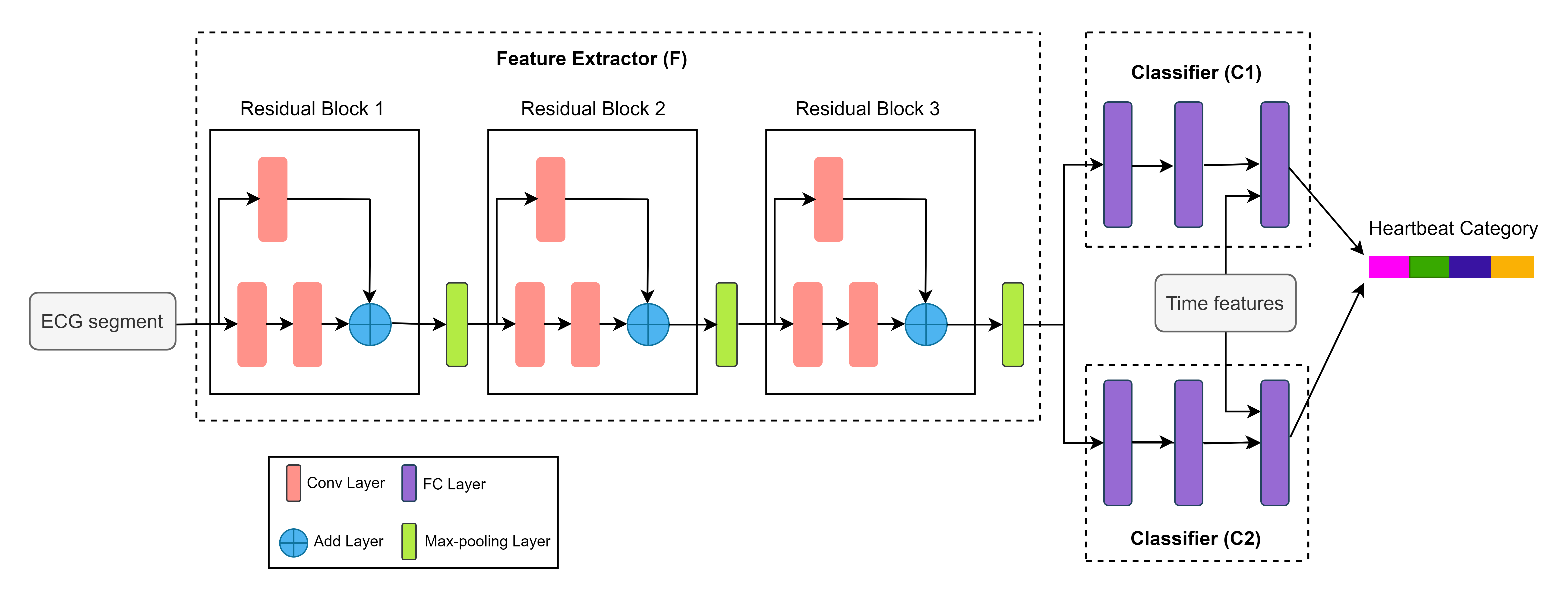}}
    \caption{Architecture of the proposed network.}
    \label{model}
\end{figure*}

The proposed method consists of the following stages.

\subsubsection{Pre-training} We train the model using the labeled source data and use a distributionally robust optimization (DRO) technique \cite{sagawa2019distributionally} during the training process. DRO enables us to train models that reduce the worst-case loss in a set of predefined groups during the training process. To prevent the model from relying on false correlations, which could lead to high losses on some data groups, we opt to train the model to reduce the worst-case loss across training data groups.

Assume predicting labels $y \in Y$ taking input $x \in X$. Let we have a model family $\Theta$ and loss $l$ : $\Theta$ × ($X$ × $Y$) $\rightarrow$ $\mathbb{R_+}$. The training samples come from a distribution $P$. The usual objective is to obtain a model $\theta \in \Theta$ that reduces the expected loss $ \mathbb{E_P} [ l ( \theta ; (x,y))]$ within the distribution $P$. The typical method for achieving this objective in training is through empirical risk minimization (ERM):

\begin{align}
\hat{\theta}_{ERM} := \underset{\theta\in \Theta}{arg min} 
 \: \mathbb{E}_{(x,y) \sim \hat{P}} [ l ( \theta ; (x,y))]
\end{align}
where $\hat{P}$ denotes the empirical distribution over the training set.
In DRO \cite{ben2013robust, duchi2021statistics, sagawa2019distributionally}, we intend to reduce the worst-case anticipated loss by considering an uncertain set of distributions $\mathlarger{\varrho}$:

\begin{align}
\underset{\theta\in \Theta}{min} \mathlarger\{ R(\theta) : =\underset{Q\in   \mathlarger {\varrho}}{sup} \: \mathbb{E}_{(x,y) \sim Q} [ l ( \theta ; (x,y))]\mathlarger\} 
\end{align}

The uncertainty set $\mathlarger{\varrho}$ represents the range of potential test distributions that our model should be capable of performing effectively on. If we select a general family $\mathlarger{\varrho}$, like a divergence ball centered around the training distribution, it can make our model more resilient to various distributional shifts. However, this approach can also result in excessively cautious models that optimize for implausible worst-case distributions.

We apply DRO to the weighted cross-entropy loss and obtain the \textit {classification loss} $L_{cls}$. Additionally, we calculate the \textit {classifier discrepancy loss} ($L_{dis}$) by measuring the Euclidean distance between outputs from the two classifiers. The ultimate objective function in the pre-training stage is the weighted sum of classification loss and classifier discrepancy loss: 

\begin{align}
L= L_{cls} + \alpha  L_{dis}
\end{align}

where $\alpha$ is a hyperparameter. Upon completion of the pre-training stage, we obtain a robust model that exhibits a high degree of precision in classifying heartbeats within the source domain.

\subsubsection{Cluster-centroid computing}

The cluster hypothesis is a basic assumption in the classification task \cite{hearst1996reexamining}. Samples that belong to the same class should reside in the same cluster, whereas those from different classes should be at a considerable distance from each other. Following this fundamental assumption, we initially determine the centroids of the clusters in the source domain by averaging the outputs of the feature extractor for every heartbeat category. Then we train the model using two loss functions: the \textit {cluster-compacting loss} (4) and the \textit {cluster-separating loss} (5) \cite{wang2021inter} along with the classification loss to simultaneously decrease the intra-class spacing and increase the inter-class spacing. 

\begin{align}
L_{comp}= \Sigma_{k=1}^{K}\Sigma_{i=1}^{n_k} \: D(\mathbb{E}[X_k], X_{k,i})
\end{align}

\begin{align}
L_{sep}= \Sigma_{k\not=l}^{K}\Sigma_{l=1}^{K} \: max(T_m-D(\mathbb{E}[F(X_l)], \mathbb{E}[F(X_k)]), 0) 
\end{align}

where $n_k$ is the number of samples belonging to the $k^{th}$ category and $K$ is the number of heartbeat categories. $D$ represents the Euclidean distance. We compute the $L_{comp}$ and  $L_{sep}$ using the cluster centroids in the source domain. The responsibility of \textit {cluster-compacting loss} is to minimize the distance between samples within a class. The \textit {cluster-separating loss} aims to keep the cluster centroids of distinct categories apart from each other in order to reach a pre-defined threshold $T_m$ \cite{wang2021inter}. We call the cluster-related loss functions together cluster loss ($L_{ctr}$). The ultimate loss function here is the weighted sum of classification loss and cluster loss: 

\begin{align}
L= L_{cls} + {\gamma}_1 L_{comp} + {\gamma}_2 L_{sep}
\end{align}

where $\gamma_1$ and $\gamma_2$ are hyperparameters. Once the model has been trained to optimize the source clusters, we compute the centroids of the well-organized clusters ($CC_s$). As we do not have labels for the target domain, we first identify confident predictions ($CP_t$) in the target domain to compute target cluster centroids. To identify the confident predictions in the target domain we propose a novel high-confident prediction technique. We first calculate the \textit {mean intra-cluster distance} (7) and the \textit {mean classifier discrepancy} (8) in the source domain. The \textit {mean intra-cluster distance} measures the average distance between the samples in a cluster and the center of that cluster. The \textit {mean classifier discrepancy} is the average difference between the outputs of the two classifiers. 

\begin{align}
M_{ctr}= \frac{1}{n_s^k} \Sigma_{i=1}^{n_s^k} \: D(F(X_i), CC_s^k) \:for \: each \: k \in K
\end{align}

\begin{align}
M_{dis}= \frac{1}{N_s} \Sigma_{i=1}^{N_s} \: D(C_{1,i}, C_{2,i})
\end{align}
where $N_s$ is the total number of samples in the source domain and $n_s^k$ is the number of samples belonging to the $k^{th}$ category.

We feed the target domain data into the pre-trained network. If the softmax value is greater than 0.99 for a particular class,  it is a candidate for confident prediction. Instead of choosing it as a confident prediction, we further verify two additional conditions. We calculate the discrepancy between the feature extractor's output for that prediction and the source cluster centroid of the predicted class and check if the difference is less than $M_{ctr}$. We also verify whether the discrepancy between the outputs of the two classifiers for that prediction is less than $M_{dis}$. If it satisfies all the conditions, we consider it a confident prediction. In some cases, the model confidently predicts the wrong class. That is why relying only on the softmax score does not suffice. This technique filters out misclassifications by the classifier and misleading features from the feature extractor, thereby decreasing the likelihood of the model making erroneous confident predictions. After obtaining the confident predictions, we calculate the cluster centroids ($CC_t$) for the target domain based on them.

\subsubsection{Domain adaptation}
This stage minimizes the gap between the source and target domains and organizes the clusters efficiently. We feed batches from both the source and target domains. We introduce two new loss functions: the \textit{inter-domain cluster discrepancy loss} (9) and the \textit{running combined loss} (10), which are added to the \textit{cluster-compacting loss} (4) and the \textit{cluster-separating loss} (5), along with the classification loss, to train the model.

\begin{align}
L_{cd}= \Sigma_{k=1}^{K}\: D(CC_s^k, CC_t^k)
\end{align}

\begin{align}
L_{cmd}= \Sigma_{k=1}^{K}\: D(CC_{m,i}^k, CC_{m}^k) \:\{for \:all \:i : \: 1<= i <= \frac{N_s}{N_b}\}
\end{align}

where $N_b$ is the number of samples in a batch and 
\begin{align}
CC_{m}^k=avg(CC_s^k, CC_t^k)
\end{align}

We calculate the \textit {cluster-compacting loss} and the \textit {cluster-separating loss} for both the source and target domains using the source and target clusters, respectively, that are computed in the cluster computing stage. The \textit {inter-domain cluster discrepancy loss} minimizes the cluster shift of the target domain from the source domain. To calculate the \textit {running combined loss}, we first calculate the average of the cluster centroids of the source and target domains, which are computed in the cluster computing stage. We call it global average cluster centroids. For each training batch, we compute the difference between the current average cluster centroids and the calculated global average cluster centroids. This provides us with how the currently predicted clusters deviate from the previously calculated standard clusters. The final loss function is calculated as a weighted sum of classification loss, cluster-compacting loss, cluster-separating loss, inter-domain cluster discrepancy loss, and running combined loss:

\begin{align}
\begin{split}
            L =L_{cls} + \beta_1 (L_{comp}^s + L_{comp}^t) + \beta_2 (L_{sep}^s + L_{sep}^t) \\+ \beta_3 L_{cd} + \beta_4 L_{cmb}
\end{split}
\end{align}

where $\beta_1$, $\beta_2$, $\beta_3$, and $\beta_4$ are hyperparameters. Algorithm 1 illustrates the entire algorithm.

\subsection{Data preprocessing and network inputs}
We use a special input format that allows for the inclusion of both morphological and temporal information. Initially, the ECG signal undergoes preprocessing to create the input data, which is then fed into the model. The classifier receives input from both the model's extracted features and the manually extracted time features, which are concatenated together to produce the ultimate classification outcome. The preprocessing consists of the following steps:

(1) Signal denoising: We first filter the ECG signal by a bandpass filter with a passband of 3-20 Hz to eliminate the influence of power-line interference, muscle artifacts, baseline wander, and electrode contact noise.

(2) Resampling: The ECG sampling frequency varies across different datasets, with the MITDB, INCARTDB, and ESTDB datasets having sampling frequencies of 360 Hz, 257 Hz, and 250 Hz, respectively. We use an FIR filter to resample the data, resulting in a unified sampling rate of 256 Hz to address the challenge of varying sampling rates.

(3) Heartbeat segmentation: We partition the ECG signal into multiple small segments, utilizing the R-peaks as a reference point. To keep the input dimension fixed, we segment the ECG signal into segments of fixed length. We first compute the RR intervals from the R-peaks for all the ECG signals belonging to the source domain. Then, we compute the arithmetic mean ($RR_{mean}$) of the RR intervals. Suppose,
$R_i$ is the R-peak position of the $i^{th}$ heartbeat. The corresponding segment starts at position $(R_i-\lfloor\frac{1}{2}RR_{mean}\rfloor)$ and ends at position $(R_i+\lfloor\frac{1}{2}RR_{mean}\rfloor)$.

(4) Time feature extraction: Taking inspiration from Niu et al. \cite{niu2020deep}, we extract three time features, including the current RR-interval ($RR_{curr}$), an average of the pre-RR intervals ($RR_{pre}$) from the beginning to the current position, and an average of the last eight pre-RR intervals ($RR_{pre8}$) from the current position.

\begin{algorithm}
\begin{algorithmic}
\REQUIRE 
\hfill
\\Source samples and labels: $X_s$, $Y_s$;

Target samples: $X_t$;

Feature extractor: $F$, classifiers: $C_1$, $C_2$;

Epochs: $E_1$, $E_2$, $E_3$; $\alpha$, $\gamma_1$, $\gamma_2$, $\beta_1$, $\beta_2$, $\beta_3$, $\beta_4$, $T_m$;

Batch size: $N_b$, source size: $N_s$, target size: $N_t$, no of heartbeat categories: $K$; $RR_{curr}$, $RR_{pre}$, $RR_{pre8}$;
\ENSURE
\hfill

  \For{$i$ = 1 to $E_1$}{
      \For{$j$ = 1 to $\frac{N_s}{N_b}$}{
       Compute classification loss $L_{cls}$:  
       
       \hspace*{4mm} Compute weighted cross entropy loss;
       \hspace*{4mm} Apply DRO technique on the loss;
       
       Compute classifier discrepancy loss $L_{dis}$;
       
       $L$ =$L_{cls}$ + $\alpha$$L_{dis}$;

       Update $F$, $C_1$, and $C_2$ with $L$;
    }
  }
Compute source cluster centroids $CC_s$:

    \hspace*{4mm} Compute mean $F(X_s^k)$ for each $k$ in $K$;

  \For{$i$ = 1 to $E_2$}{
      \For{$j$ = 1 to $\frac{N_s}{N_b}$}{
            Compute classification loss $L_{cls}$;

            Compute cluster-compacting loss:
            
                \hspace*{4mm} $L_{comp}$= $\Sigma_{k=1}^{K}$$\Sigma_{t=1}^{n_s^k}$ $D(F(X_s^{k,t}),CC_s^k)$; 

            Compute cluster-separating loss:
            
                \hspace*{4mm} $L_{sep}$= $\Sigma_{k\not=l}^{K}$$\Sigma_{l=1}^{K}$max($T_m$-$D(CC_s^k,CC_s^l)$,0); 
            
           \hspace*{4mm} $L$ =$L_{cls}$ + $\gamma_1$$L_{comp}$ + $\gamma_2$$L_{sep}$;

           Update $F$, $C_1$, and $C_2$ with $L$; 
    }
  }

Compute source cluster centroids $CC_s$; 

Compute mean intra-cluster distance $M_{ctr}$:

   \hspace*{4mm} Compute mean $D(F(X_s^k),CC_s^k)$ for each $k$ in $K$;
   
Compute mean classifier discrepancy $M_{dis}$:

   \hspace*{4mm} Compute mean $D(C_1,C_2)$ from all source samples;

Compute target cluster centroids $CC_t$:

    \hspace*{4mm} Pick confident predictions $CP_t$:
    
    \hspace*{6mm} For $i^{th}$ sample ($X_t^i$) in target domain;
    
    \hspace*{8mm} \If{softmax score $>$ 0.99 for $k^{th}$ class}{

            \hspace*{10mm} \If{$D(F(X_t^i),CC_s^k)$ $<$ $M_{ctr}^k$ and \hspace*{14mm}$D(C_1^i,C_2^i)$ $<$ $M_{dis}$}{

                \hspace*{12mm} Pick it as a confident prediction;
            }
        
   }
   \hspace*{4mm} Compute target cluster centroids $CC_t$ using $CP_t$;

\For{$i$ = 1 to $E_3$}{
    \For{$j$ = 1 to $\frac{N_s}{N_b}$}{
            Compute classification loss $L_{cls}$ for source; 
            
            Compute compacting and separating loss for source and target $L_{comp}^s$, $L_{comp}^t$, $L_{sep}^s$, $L_{sep}^t$;

            Compute inter-domain cluster discrepancy $L_{cd}$;

            Compute running combined loss $L_{cmb}$;
            
            $L$ =$L_{cls}$ + $\beta_1$ ($L_{comp}^s$ + $L_{comp}^t$) + 
           
           \hspace*{5mm} $\beta_2$ ($L_{sep}^s$ + $L_{sep}^t$) + $\beta_3$$L_{cd}$ + $\beta_4$$L_{cmb}$;
            
            Update $F$, $C_1$, and $C_2$ with $L$; 
        }
   }
\caption{Unsupervised domain adaptation.}
\end{algorithmic}
\end{algorithm}
\section{Experiments}
\label{sec:experiments}
\subsection{Databases}
The proposed model has been trained with the MIT-BIH Arrhythmia Database (MITDB) and tested using the St. Petersburg INCART 12-lead Arrhythmia Database (INCARTDB) and the European ST-T Dataset (ESTDB). These are public databases that have been widely utilized for testing in arrhythmia heartbeat classification tasks. Both the cross-database and cross-channel paradigms are taken into account in this study. According to the American National Standards Institute/Association for the Advancement of Medical Instrumentation (ANSI/AAMI EC57: 1998) standard, every heartbeat can be categorized into one of five groups: Normal heartbeat (N), ventricular ectopic heartbeat (V), supraventricular ectopic heartbeat (S), fusion heartbeat (F), and unknown heartbeat (Q). This study investigates classifying four types of heartbeats: N, V, S, and F, as the Q-type data is inadequate in size and cannot be utilized as a reliable basis for evaluating the classification results.
\subsubsection{MIT-BIH Arrhythmia Database (MITDB)}
The MIT-BIH Arrhythmia Database contains 48 ECG recordings obtained from 47 subjects. Each record has two-channel signals of 30 minutes sampled at 360 Hz. Our training set excludes 4 paced records (102, 104, 107, 217), as per the ANSI/AAMI convention. As the normal QRS complex of ML II is typically prominent in the MIT-BIH database, this experiment only uses 44 ML II recordings. 

\subsubsection{St. Petersburg INCART 12-lead Arrhythmia Database (INCARTDB)}

The St. Petersburg INCART 12-lead Arrhythmia Database contains 75 ECG recordings with 12 standard leads. Each recording is 30 minutes long and sampled at 257 Hz.  

\subsubsection{European ST-T Dataset (ESTDB)}
The European ST-T Dataset consists of 90 records from 79 subjects. Each recording lasts two hours and is sampled at 250 Hz. 

\subsection{Experimental setting}
The implementation of our method is carried out using the PyTorch framework. The optimization of the model parameters is achieved by utilizing the Adam optimizer, which is a widely-used stochastic gradient descent algorithm. The classification loss is measured using the weighted cross-entropy loss function. We set the batch size to 512 and the learning rate and weight decay to 0.001 and 0.0005, respectively. The number of arrhythmic beats is significantly lower compared to the number of normal beats in all three databases. For example, the percentage of normal, ventricular ectopic, supraventricular ectopic, and fusion beats in the MIT-BIH Arrhythmia Database is 89.48\%, 6.96\%, 2.76\%, and 0.80\%, respectively. To reduce the existing imbalances, we duplicate the data for ventricular ectopic, supraventricular ectopic, and fusion beats by factors of 2, 5, and 10, respectively, and incorporate them into the datasets. The augmentation by the same factor is performed for all three databases. Table I shows the number of ECG records, ECG segments, and samples for each of the heartbeat categories in all databases before and after augmentation.

\begin{table*}
\caption {Number of samples in the dataset}
\label{table1}
\centering
\begin{tabular}{c p{0.1\linewidth} p{0.1\linewidth} p{0.1\linewidth} p{0.1\linewidth} p{0.1\linewidth} p{0.1\linewidth} p{0.1\linewidth} }
\hline
  & \multicolumn{5}{c}{MITDB} &\\
\hline

 & Records & Segments & N & V & S & F \\
\hline
Before augmentation & 44 & 100718 & 90125 & 7009 & 2781 & 803\\

After augmentation & 44 & 136671 & 90125 & 21027 & 16686 & 8833 \\
\hline
  & \multicolumn{5}{c}{} &\\
  & \multicolumn{5}{c}{INCARTDB} &\\
\hline

Before augmentation & 75 & 175868 & 153676 & 20013 & 1960 & 219\\

After augmentation & 75 & 227884 & 153676 & 60039 & 11760 & 2409 \\
\hline
  & \multicolumn{5}{c}{} &\\
  & \multicolumn{5}{c}{ESTDB} &\\
\hline

Before augmentation & 90 & 790549 & 784633 & 4467 & 1095 & 354\\

After augmentation & 90 & 808498 & 784633 & 13401 & 6570 & 3894 \\
\hline
\end{tabular}\\

\end{table*}

\subsection{Results and discussion}

To evaluate the efficacy of our proposed approach, we first evaluate it against a method that has an identical network architecture and the same experimental settings as our proposed method but excludes the domain-adaptation aspect. Next, we evaluate five recent approaches \cite{sun2016deep,sagawa2019distributionally,huang2020self,wang2021inter,niu2020deep} known for their high performance using the same network architecture and experimental setting as our proposed method. We compare our proposed method against these approaches. Moreover, we perform an ablation analysis to demonstrate the impact of the individual components of our proposed approach.

We train our model using the lead II MITDB records and test our method on three datasets: DS 1, DS 2, and DS 3. DS 1 is used to test the model using a cross-database paradigm, where the source and target data are from different databases but come from the same channel. Lead II records from the INCARTDB constitute DS 1. DS 2 is used to test the model using a cross-database and cross-channel paradigm, where the source and target data come from different databases and channels. DS 2 consists of lead V5 INCARTDB records. Lead V5 records from the ESTDB make up DS 3. Although some recent domain-adaptation-based methods report results on ECG classification \cite{niu2020deep,jin2022multi,hasani2020classification}, they use different groups as source-target within the same database or use non-public databases. None of them employ the same train-test configuration as this investigation. 

Table II presents a comparison of the overall accuracy of two methods: our proposed domain-adaptive method and the method that has the same network architecture and experimental settings as ours but does not include the domain-adaptation aspect. Our method improves the overall accuracy by 14.24\%, 10.89\%, and 10.21\% on DS 1, DS 2, and DS 3, respectively. We show the average results of five trials, as every trial uses random training and test batches. The comparison results in terms of sensitivity (Se), positive predictive value (PPV), and F1 score are shown in Table III.  Although the method without domain adaptation yields satisfactory results for normal heartbeats, it encounters major difficulties when dealing with arrhythmic beats. Our method achieves a significant increase in performance and reaches the best F1 scores on ventricular ectopic (79.66\%), supraventricular ectopic (46.87\%), and fusion (13.96\%) compared to 11.15\%, 0.51\%, and 4.74\%, respectively, achieved by the method without domain adaptation. From the confusion matrices (Fig. 3), we can see that although the method excluding domain adaptation identifies normal heartbeats satisfactorily, it identifies most of the arrhythmic heartbeats as normal beats. It identifies more than 99\% of supraventricular ectopic beats and over 88\% of fusion beats as normal beats in all three test datasets. Our proposed method shows a notable performance improvement in this particular scenario.

\begin{table*}
\caption {Overall accuracy comparisons between the proposed method and the method without domain adaptation}
\label{table2}
\centering
\begin{tabular}
{ p{0.19\linewidth} p{0.17\linewidth} p{0.17\linewidth} p{0.06\linewidth} }
\hline
  & \multicolumn{2}{c}{Overall accuracy (\%)} &\\
\hline

 & INCARTDB & INCARTDB & ESTDB \\
 &(cross-domain paradigm)& (cross-domain and cross-channel paradigm)&\\
\hline
Without domain adaptation & 70.37 & 71.43 & 66.23\\

Proposed method & \textbf{84.61} & \textbf{82.32} & \textbf{76.44}\\
\hline
\end{tabular}\\
\end{table*}

\begin{table*}
\caption {Comparison of the results between the proposed method and the method without domain adaptation}
\label{table3}
\centering
\begin{tabular}
{p{0.17\linewidth} p{0.043\linewidth} p{0.056\linewidth}  p{0.044\linewidth} p{0.043\linewidth} p{0.056\linewidth} p{0.044\linewidth} p{0.043\linewidth} p{0.056\linewidth} p{0.044\linewidth} p{0.043\linewidth} p{0.056\linewidth} p{0.044\linewidth}  }
\hline
 & \multicolumn{11}{c}{INCARTDB (cross-domain paradigm)} &\\
\hline
 &  & N & & & V & & & S & & & F & \\
\hline
& Se (\%)& PPV (\%)& F1 (\%) & Se (\%) & PPV (\%) & F1 (\%) & Se (\%) & PPV (\%) & F1 (\%) & Se (\%) & PPV (\%) & F1 (\%)\\
\hline
Without domain adaptation & 96.04 & 74.32 & 83.79 &  2.67 &  97.10 & 5.20 & 0.32 &  0.98 &  0.48 &  3.33 &  1.38 & 1.95\\

Proposed method & 91.08 & 92.27 & \textbf{91.67} & 72.98 & 87.69 & \textbf{79.66} & 47.67 & 40.65 & \textbf{43.88} & 33.77 & 8.80 & \textbf{13.96}\\
\hline
& \multicolumn{11}{c}{} &\\
 & \multicolumn{11}{c}{INCARTDB (cross-domain and cross-channel paradigm)} &\\
\hline

Without domain adaptation & 97.58 & 73.80 & 84.04 & 2.83 & 94.59 & 5.49 & 0.0 & 0.0 & 0.0 & 7.77 & 3.41 & 4.74\\

Proposed method & 86.38 & 92.11 & \textbf{89.15} & 77.90 & 71.13 & \textbf{74.36} & 58.79 & 38.97 & \textbf{46.87} & 6.71 & 7.09 & \textbf{6.90}\\
\hline

& \multicolumn{11}{c}{} &\\
 & \multicolumn{11}{c}{ESTDB} &\\
\hline

Without domain adaptation & 98.31 & 67.88 & 80.31 & 5.92 & 95.75 & 11.15 & 0.30 & 1.65 & 0.51 & 5.88 & 3.28 & 4.21\\

Proposed method & 89.61 & 81.12 & \textbf{85.16} & 65.28 & 95.02 & \textbf{77.39} & 24.84 & 19.59 & \textbf{21.90} & 10.77 & 5.36 & \textbf{7.16}\\
\hline
\end{tabular}\\
\end{table*}

\begin{figure*}[!t]
     \centering
     \begin{subfigure}[b]{0.45\textwidth}
         \centering
         \includegraphics[width=\textwidth]{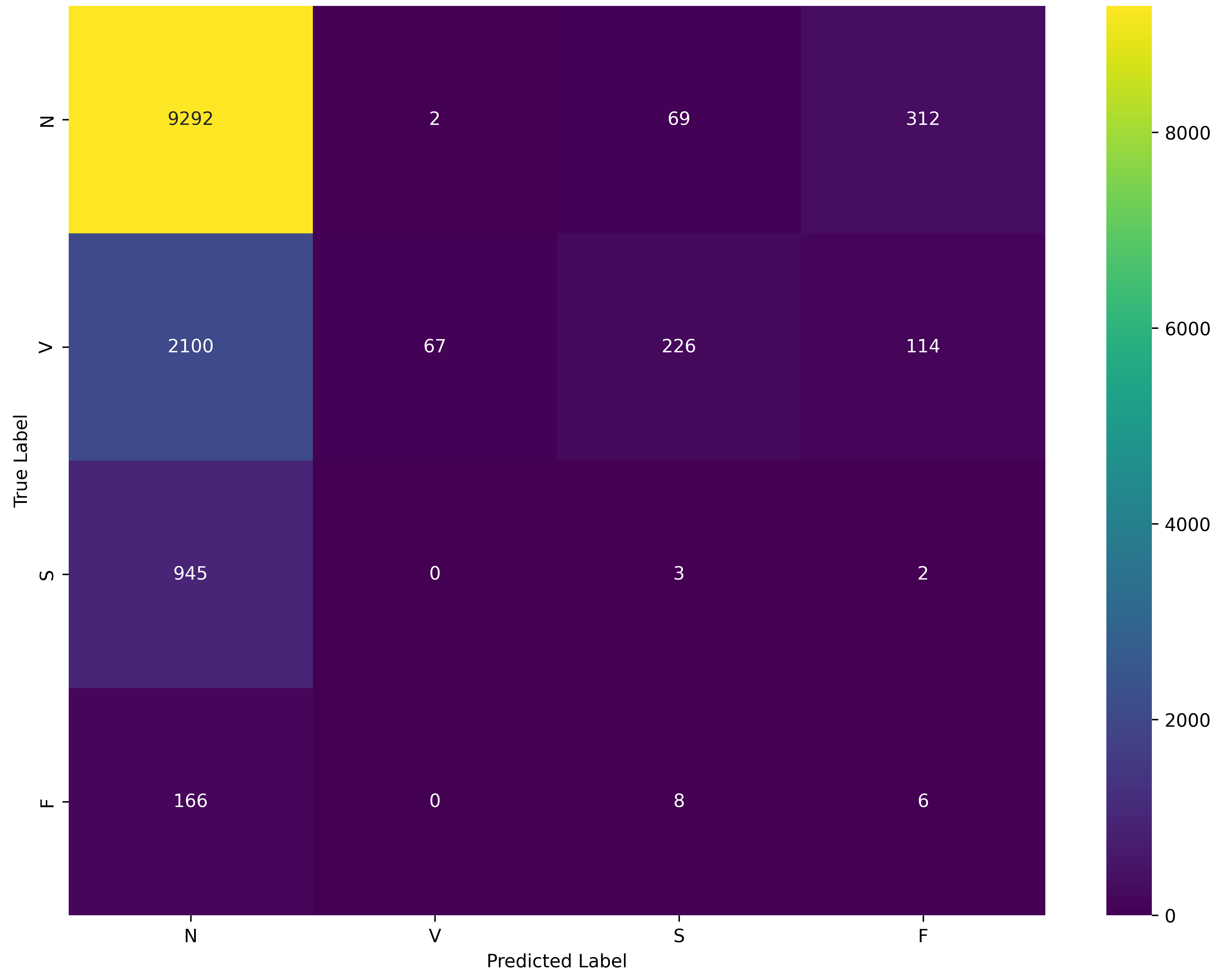}
         \caption{}
         \label{fig:a}
     \end{subfigure}
     \hfill
     \begin{subfigure}[b]{0.45\textwidth}
         \centering
         \includegraphics[width=\textwidth]{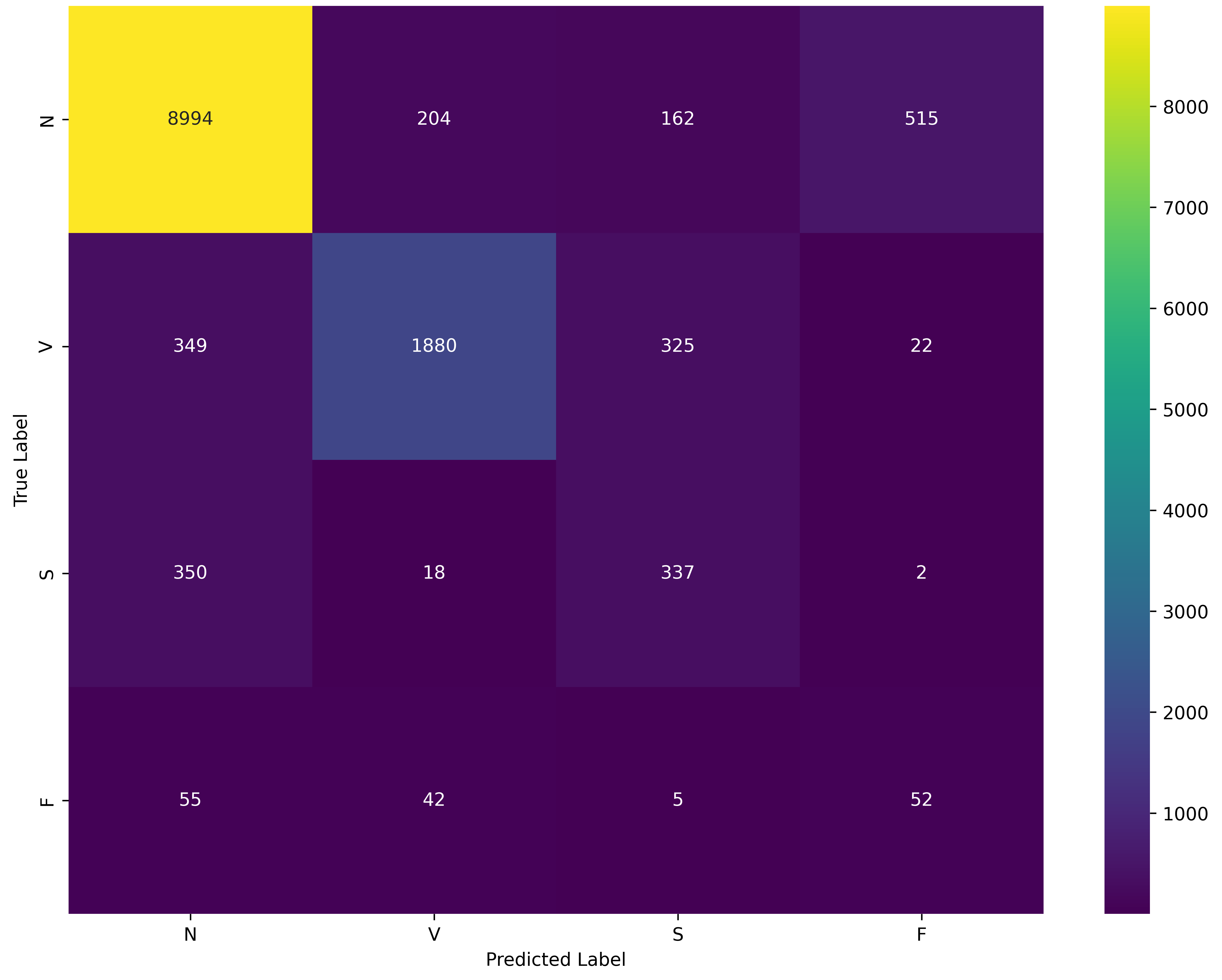}
         \caption{}
         \label{fig:b}
     \end{subfigure}
     \hfill
         \begin{subfigure}[b]{0.45\textwidth}
         \centering
         \includegraphics[width=\textwidth]{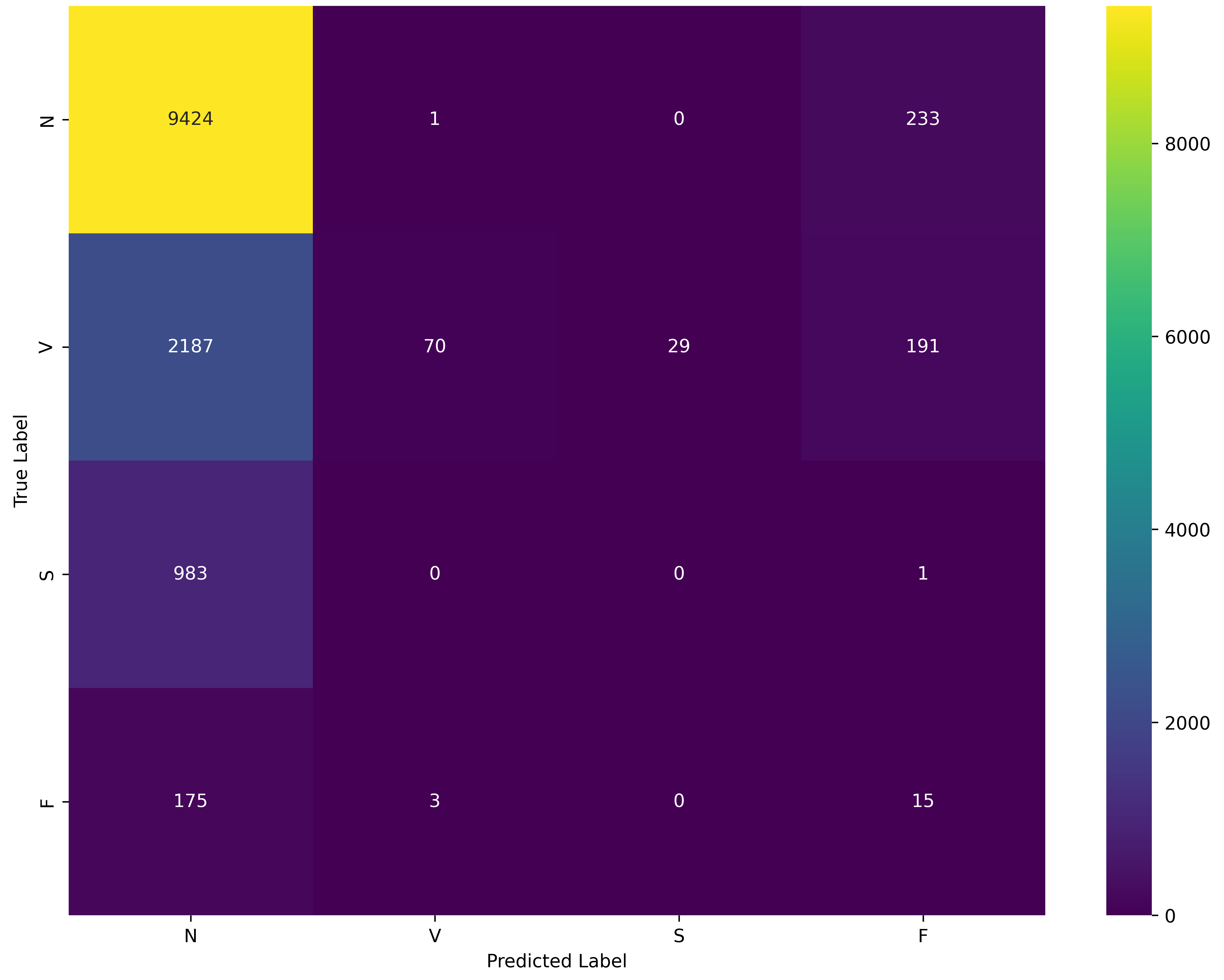}
         \caption{}
         \label{fig:c}
     \end{subfigure}
    \hfill
             \begin{subfigure}[b]{0.45\textwidth}
         \centering
         \includegraphics[width=\textwidth]{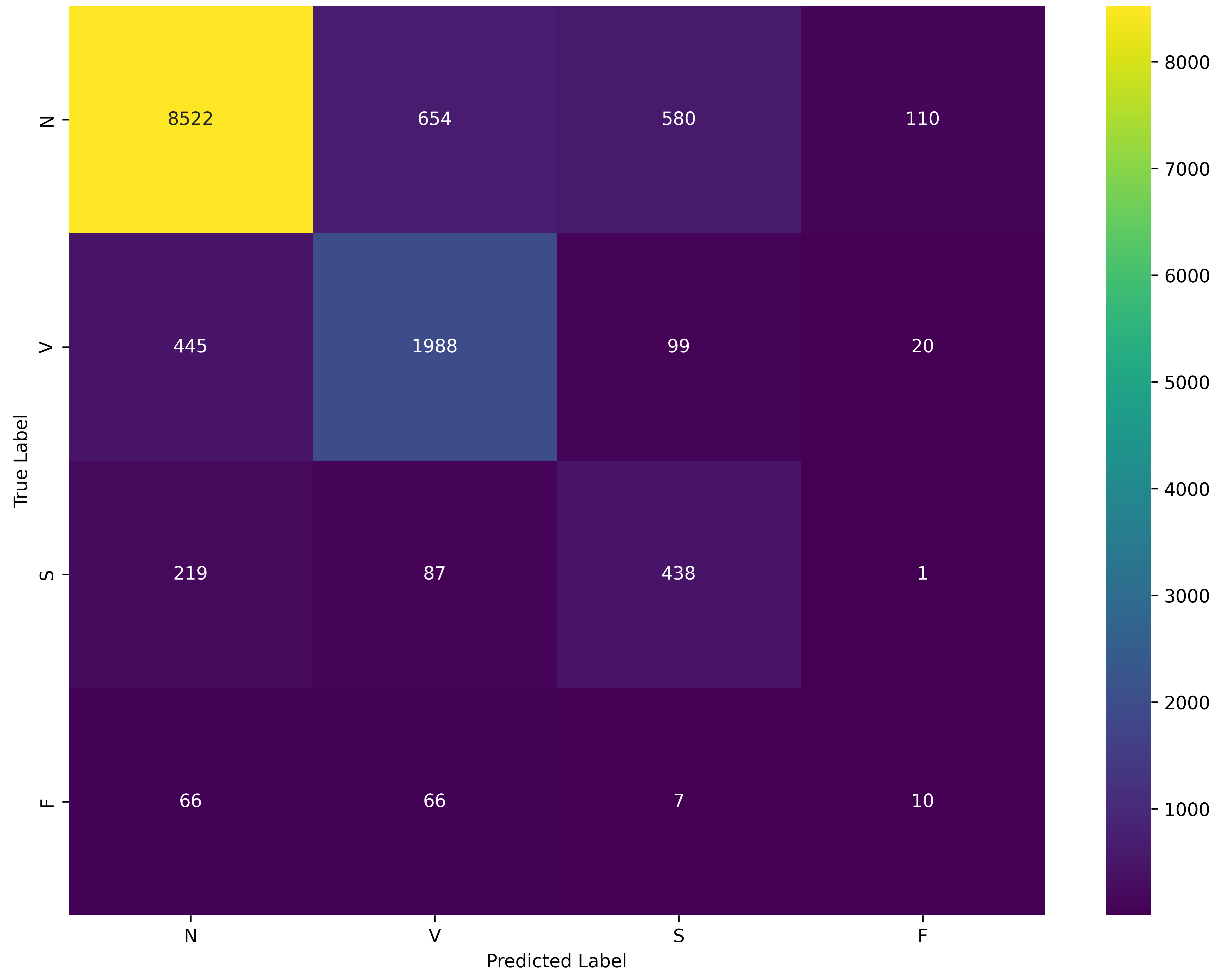}
         \caption{}
         \label{fig:d}
     \end{subfigure}
     \hfill
     \begin{subfigure}[b]{0.45\textwidth}
         \centering
         \includegraphics[width=\textwidth]{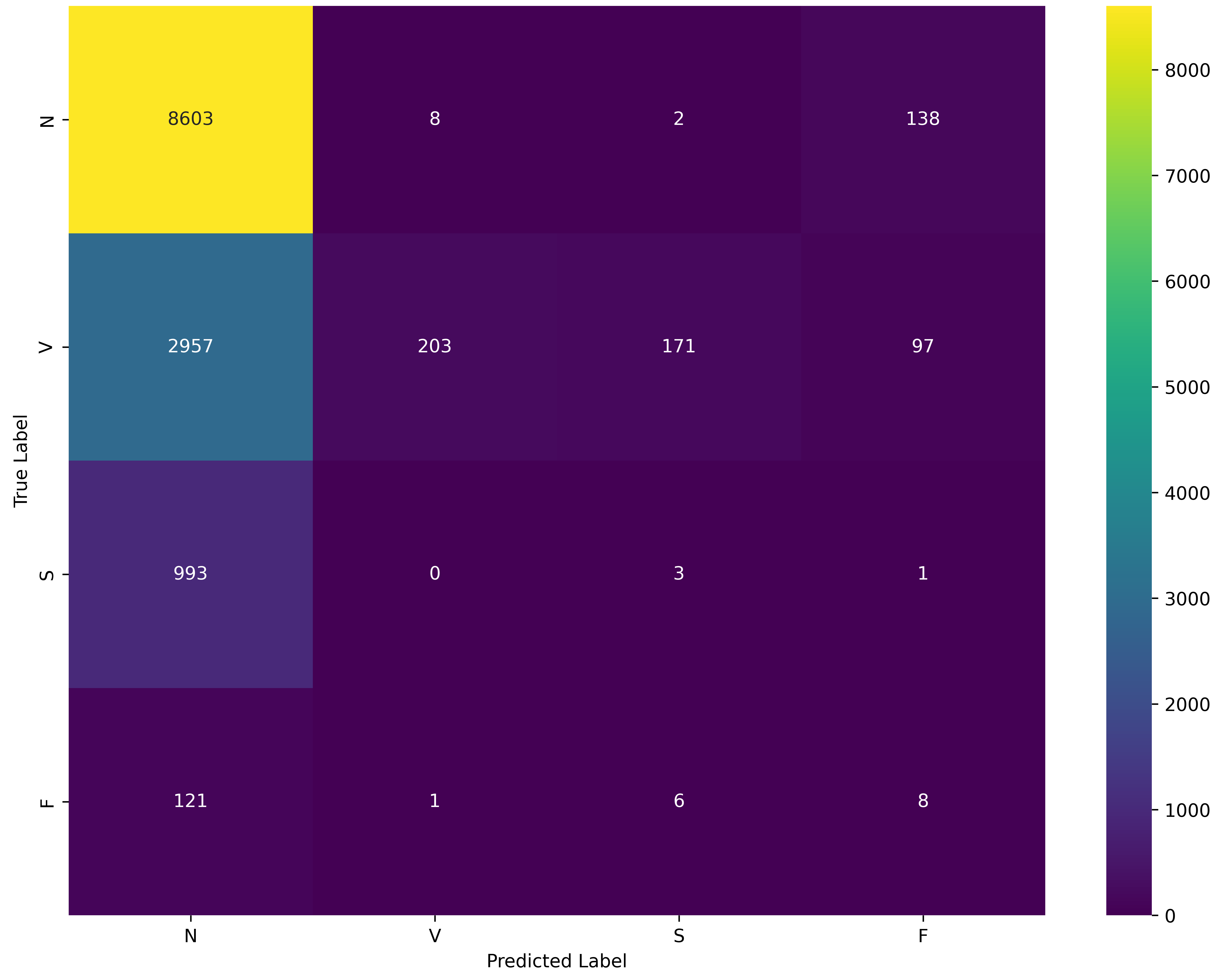}
         \caption{}
         \label{fig:e}
     \end{subfigure}
     \hfill
         \begin{subfigure}[b]{0.45\textwidth}
         \centering
         \includegraphics[width=\textwidth]{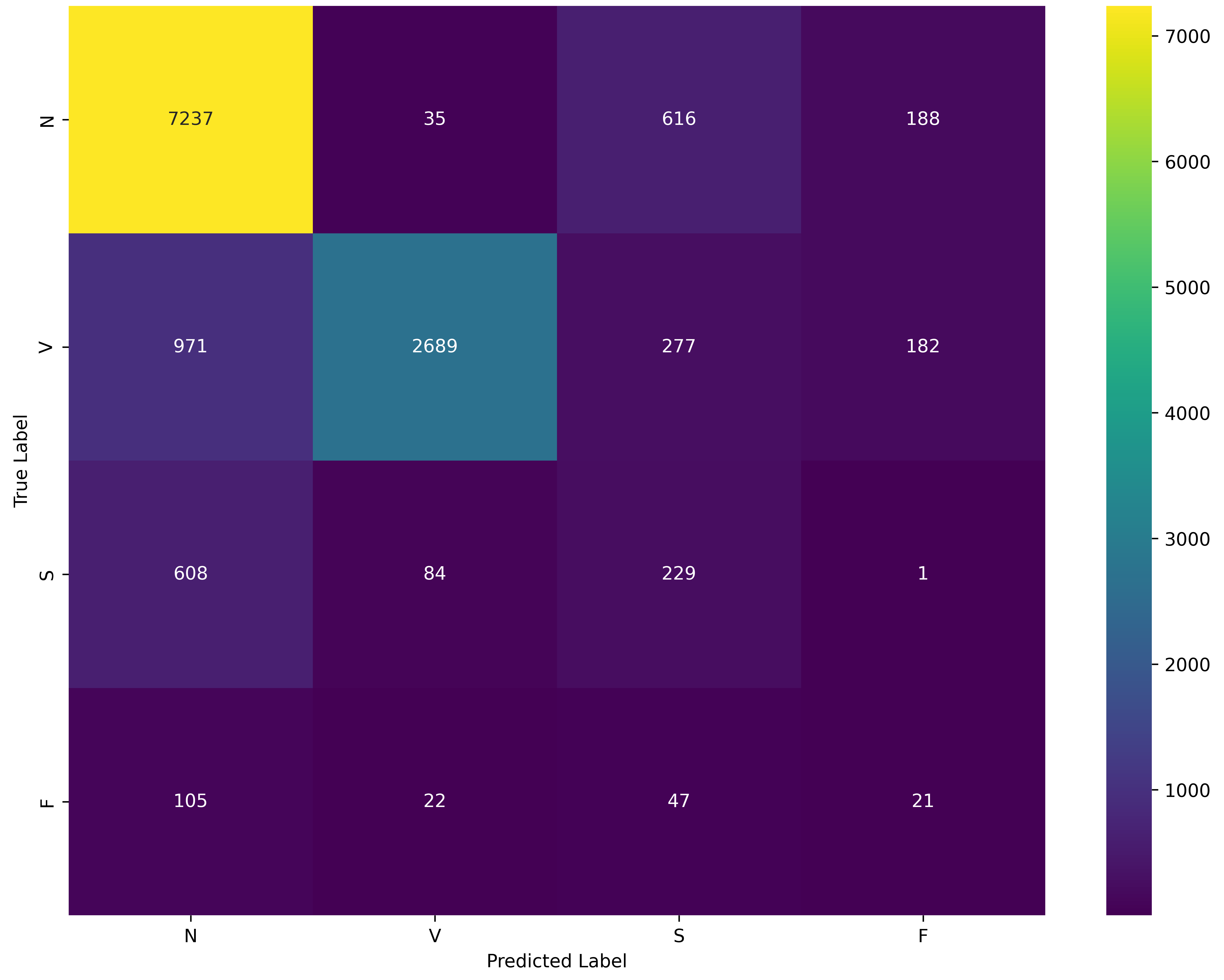}
         \caption{}
         \label{fig:f}
     \end{subfigure}
        \caption{Confusion matrices: Figures (a) and (b) show the confusion matrices for the method without domain adaptation and our proposed method, respectively, on the INCARTDB dataset (cross-domain paradigm). Figures (c) and (d) show the confusion matrices for the method without domain adaptation and our proposed method, respectively, on the INCARTDB dataset (cross-domain and cross-channel paradigm). Figures (e) and (f) show the confusion matrices for the method without domain adaptation and our proposed method, respectively, on the ESTDB dataset. }
        \label{fig:six matrices}
\end{figure*}

Overall, we observe a minor enhancement in the detection of normal heartbeats, but a substantial improvement in the case of detecting arrhythmic heartbeats over the method without domain adaptation. In Table I, we see that although we increase the number of arrhythmic beats of all types, they still remain significantly lower compared to normal beats. As an illustration, out of the total number of beats, the percentage of fusion beats is only 6.46\% in MITDB, 1.06\% in INCARTDB, and 0.48\% in ESTDB. The model finds it challenging to achieve satisfactory performance for arrhythmic heartbeats due to the considerably low number of samples available.

We employ the same network architecture and experimental setup to evaluate the performance of five recent approaches that are recognized for their high accuracy \cite{sun2016deep,sagawa2019distributionally,huang2020self,wang2021inter,niu2020deep}. They worked with different types of data and different types of applications than ours, but they achieved satisfactory performances. By utilizing the data and model employed in this experiment, we evaluate the results of those approaches against our proposed method. The comparisons of overall accuracy across the three datasets are demonstrated in Table IV. Our method outperforms other approaches and achieves the highest overall accuracy of 84.61\%, 82.32\%, and 76.44\% on DS 1, DS 2, and DS 3, respectively. This is 8.53\%, 4.17\%, and 6.03\%, respectively, higher than the second-best approach.

Table V illustrates that our proposed method exhibits notable superiority in terms of sensitivity, PPV, and F1 score over all other approaches on the INCARTDB database in a cross-domain scenario. Although the method proposed by Sagawa et al. \cite{sagawa2019distributionally} achieves the highest F1 score of 92.47\% (0.8\% higher than our method) in detecting normal beats, our method performs significantly better in detecting arrhythmic beats. Missing even one arrhythmic beat can have catastrophic consequences, but identifying arrhythmias at an early stage and providing appropriate treatment can prevent heart failure.

\begin{table*}
\caption {Overall accuracy comparisons of the proposed method with other methods}
\label{table4}
\centering
\begin{tabular}
{ p{0.19\linewidth} p{0.17\linewidth} p{0.17\linewidth} p{0.06\linewidth} }
\hline
  & \multicolumn{2}{c}{Overall accuracy (\%)} &\\
\hline

 & INCARTDB & INCARTDB & ESTDB \\
 &(cross-domain paradigm)& (cross-domain and cross-channel paradigm)&\\
\hline
Sun and Saenko \cite{sun2016deep}  & 72.48 & 71.66 & 64.73\\
Sagawa et al. \cite{sagawa2019distributionally}  & 74.22 & 75.96 & 68.84\\
Huang et al. \cite{huang2020self} & 73.76 & 76.89 & 69.01\\
Wang et al. \cite{wang2021inter}   & 76.08 & 78.15 & 70.41\\
Niu et al. \cite{niu2020deep}  & 70.89 & 68.92 & 69.56\\
Proposed method & \textbf{84.61} & \textbf{82.32} & \textbf{76.44}\\
\hline
\end{tabular}\\
\end{table*}

\begin{table*}
\caption {Performance comparisons of the proposed method with other methods on INCARTDB (cross-domain paradigm)}
\label{table5}
\centering
\begin{tabular}
{p{0.13\linewidth} p{0.05\linewidth} p{0.056\linewidth}  p{0.05\linewidth} p{0.05\linewidth} p{0.056\linewidth} p{0.05\linewidth} p{0.05\linewidth} p{0.056\linewidth} p{0.05\linewidth} p{0.05\linewidth} p{0.056\linewidth} p{0.05\linewidth}  }
\hline

 &  & N & & & V & & & S & & & F & \\
\cline{2-13}
& Se (\%)& PPV (\%)& F1 (\%) & Se (\%) & PPV (\%) & F1 (\%) & Se (\%) & PPV (\%) & F1 (\%) & Se (\%) & PPV (\%) & F1 (\%)\\
\hline
Sun and Saenko \cite{sun2016deep} & 91.56 & 81.68 & 86.34 & 23.68 & 98.23 & 38.16 & 0.83 & 1.83 & 1.14 & 9.15 & 1.13 & 2.02\\

Sagawa et al. \cite{sagawa2019distributionally} & 91.62 & 93.34 & \textbf{92.47} & 27.21 & 69.52 & 39.11 & 18.98 & 10.32 & 13.37 & 4.61 & 0.54 & 0.96\\

Huang et al. \cite{huang2020self} & 96.12 & 76.35 & 85.10 & 10.84 & 58.18 & 18.27 & 8.66 & 19.53 & 12.00 & 4.08 & 5.13 & 4.55\\

Wang et al. \cite{wang2021inter} & 95.11 & 79.51 & 86.61 & 43.42 & 74.65 & 54.91 & 29.64 & 42.80 & 35.02 & 4.65 & 5.45 & 5.02\\

Niu et al. \cite{niu2020deep} & 94.93 & 76.26 & 84.58 & 1.23 & 99.81 & 2.43 & 1.75 & 2.78 & 2.15 & 5.70 & 1.77 & 2.70\\

Proposed method & 91.08 & 92.27 & 91.67 & 72.98 & 87.69 & \textbf{79.66} & 47.67 & 40.65 & \textbf{43.88} & 33.77 & 8.80 & \textbf{13.96} \\
\hline
\end{tabular}\\
\end{table*}

\subsubsection{Ablation study}

An ablation study is shown in Table VI and Fig. 4. We remove the components of our proposed method, which are the key contributions of this study, one at a time, and evaluate the model to observe the impact of each component. We form four models (Model A, Model B, Model C, Model D) by removing one component at a time while keeping everything else the same. Model A is formed by excluding the two-stage training (cluster-centroid computation and adaptation). Model B eliminates the technique of selecting confident predictions in the cluster centroid computation stage. Model C removes the inclusion of two new objective functions (\textit{inter-domain cluster discrepancy loss} and \textit{running combined loss}) in the adaptation stage. Model D excludes the inclusion of three significant time features ($RR_{curr}$, $RR_{pre}$, $RR_{pre8}$) while maintaining all other aspects unchanged. 

Each of the four modules in the proposed model has an impact on the performance to some extent, as evident from Fig. 4 and Table VI.  However, this impact varies across different modules. The two-stage training has a greater impact on performance than the others. Fig. 4 shows that eliminating two-stage training (Model A) leads to a significant reduction in performance, with an average overall accuracy decrease of 8.92\% across the three datasets. If we remove the technique of selecting confident predictions (Model B), the average overall accuracy drops by 5.94\%. The removal of the two new objective functions (Model C) results in an average overall accuracy reduction of 4.35\%. Excluding time features (Model D) leads to an average overall accuracy decrease of 3.33\% across the three datasets. Two-stage training leads to a greater enhancement in sensitivity, PPV, and F1 score among the four components (Table VI). The enhancement is more in detecting arrhythmic beats than normal beats.  It increases the F1 score for detecting ventricular ectopic, supraventricular ectopic, and fusion beats by 23.32\%, 25.24\%, and 13.96\%, respectively, on the INCARTDB database in a cross-domain paradigm. The contributions of the confident prediction selection technique, the two new objective functions, and the three time features in improving the performance of detecting arrhythmic beats are also significant. The technique of selecting confident predictions is used during the cluster-centroid computing stage, while two new objective functions are employed during the domain adaptation stage of the two-stage training. The cluster centroids obtained through the confidence prediction selection technique are utilized to compute the objective functions. Together, they create well-distinguishable clusters and minimize the difference between the source and target distributions. As a result, removing any of the three components leads to a significant drop in performance. Furthermore, the concatenation of significant time features with the deep features results in an increased feature diversity, ultimately leading to more accurate predictions by the model.

\begin{figure}[!t]
    
    \centerline{\includegraphics[width=0.4\textwidth]{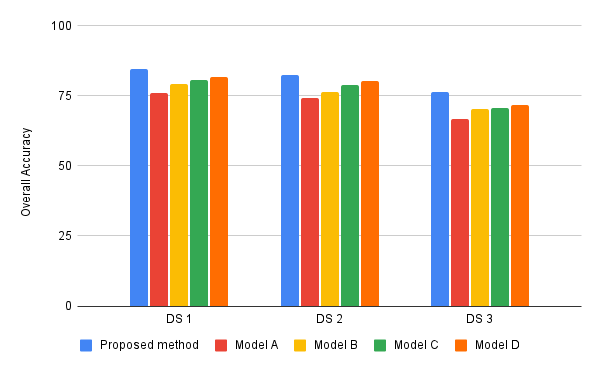}}
    \caption{Comparisons of overall accuracy on test datasets (DS 1, DS 2, and DS 3) through the ablation study on the components of the proposed method.} 
    \label{ablation}
\end{figure}

\begin{table*}
\caption {Ablation study on the components of the proposed method on INCARTDB (cross-domain paradigm)}
\label{table6}
\centering
\begin{tabular}
{p{0.11\linewidth} p{0.05\linewidth} p{0.056\linewidth}  p{0.05\linewidth} p{0.05\linewidth} p{0.056\linewidth} p{0.05\linewidth} p{0.05\linewidth} p{0.056\linewidth} p{0.05\linewidth} p{0.05\linewidth} p{0.056\linewidth} p{0.05\linewidth}  }
\hline
 &  & N & & & V & & & S & & & F & \\
\cline{2-13}
& Se (\%)& PPV (\%)& F1 (\%) & Se (\%) & PPV (\%) & F1 (\%) & Se (\%) & PPV (\%) & F1 (\%) & Se (\%) & PPV (\%) & F1 (\%)\\
\hline
Model A & 89.00 & 82.10 & 85.41 & 46.15 & 72.31 & 56.34 & 20.73 & 16.93 & 18.64 & 0.0 & 0.0 & 0.0\\

Model B & 90.56 & 87.29 & 88.90 & 53.03 & 84.97 & 65.30 & 35.52 & 27.14 & 30.77 & 10.88 & 2.55 & 4.13\\

Model C & 90.16 & 89.04 & 89.60 & 61.40 & 80.44 & 69.64 & 36.79 & 25.59 & 30.19 & 15.43 & 8.33 & 10.82\\

Model D & 87.70 & 91.21 & 89.42 & 72.95 & 87.35 & 79.50 & 43.44 & 27.11 & 33.39 & 14.37 & 4.71 & 7.09\\

Proposed method & 91.08 & 92.27 & \textbf{91.67} & 72.98 & 87.69 & \textbf{79.66} & 47.67 & 40.65 & \textbf{43.88} & 33.77 & 8.80 & \textbf{13.96}\\
\hline
\end{tabular}\\
\end{table*}

\subsubsection{Effect of hyperparmeters}
In our study, we examine the influence of hyperparameters by changing their values between 0 and 1. Fig. 5 displays the overall accuracy of our method on INCARTDB (cross-domain paradigm) for different values of the hyperparameters.  We choose the values of the hyperparameters $\alpha$, $\gamma_1$, $\gamma_2$, $\beta_1$, $\beta_2$, $\beta_3$, $\beta_4$ to be 0.5, 0.1, 0.1, 0.1, 0.1, 0.5, and 0.1, respectively, as they yield the best results for our method. The same hyperparameter values are used for all datasets (DS 1, DS 2, and DS 3) in our experiments.

\begin{figure}[!t]
    \centerline{\includegraphics[width=0.4\textwidth]{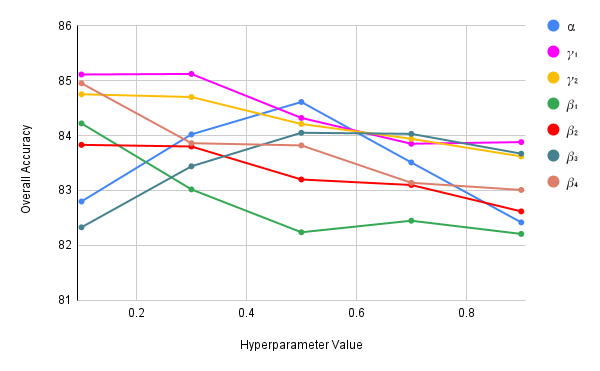}}
    \caption{The impact of hyperparameters on the performance of the model on INCARTDB (cross-domain paradigm).} 
    \label{hyperparameter}
\end{figure}

The overall findings demonstrate that the proposed method achieves satisfactory performance when compared to other methods across all three test datasets. It is worth noting, however, that the method's performance on fusion beats is limited by the excessively low number of samples. Nonetheless, our method exhibits a notable performance improvement in detecting arrhythmic beats when compared to other approaches. Furthermore, the experiments conducted in the cross-domain and cross-channel paradigms reveal that our method performs well across various databases and different ECG channels.

\section{Conclusion}
\label{sec:conclusion}
This paper proposes a novel method for classifying ECG arrhythmias that effectively addresses the problem of insufficiently labeled training samples and data distribution shifts across different domains.  We design a model based on residual networks and a bi-classifier to achieve results that are comparable to other state-of-the-art models while maintaining a better-balanced performance across various categories. To minimize distribution disparities across domains, we introduce a cluster optimization method that incorporates four distinct objective functions. We propose a novel technique for selecting confident predictions in the unlabeled target domain, which ultimately enhances the precision of separating clusters. Moreover, incorporating three significant time features into the final classifier layer and applying distributionally robust optimization during pre-training improves the model's ability to classify ECG arrhythmias. The proposed approach obviates the need for any annotations for new records and does not entail the introduction of any supplementary computational or storage resources during the inference phase. Our method exhibits the potential to significantly improve the efficacy of deep learning models in various domains and can be readily adapted to unseen data.

Our approach still has some limitations, such as when there is a significantly inadequate collection of unlabeled samples, the long-tail effect may prevent the successful improvement of minor category performance. We plan to investigate the issue in our future work. 

\bibliographystyle{IEEEtran}
\bibliography{ref} 

\begin{thebibliography}{10}
\providecommand{\url}[1]{#1}
\csname url@samestyle\endcsname
\providecommand{\newblock}{\relax}
\providecommand{\bibinfo}[2]{#2}
\providecommand{\BIBentrySTDinterwordspacing}{\spaceskip=0pt\relax}
\providecommand{\BIBentryALTinterwordstretchfactor}{4}
\providecommand{\BIBentryALTinterwordspacing}{\spaceskip=\fontdimen2\font plus
\BIBentryALTinterwordstretchfactor\fontdimen3\font minus
  \fontdimen4\font\relax}
\providecommand{\BIBforeignlanguage}[2]{{%
\expandafter\ifx\csname l@#1\endcsname\relax
\typeout{** WARNING: IEEEtran.bst: No hyphenation pattern has been}%
\typeout{** loaded for the language `#1'. Using the pattern for}%
\typeout{** the default language instead.}%
\else
\language=\csname l@#1\endcsname
\fi
#2}}
\providecommand{\BIBdecl}{\relax}
\BIBdecl

\bibitem{mendis2011global}
S.~Mendis, P.~Puska, B.~e. Norrving, W.~H. Organization \emph{et~al.},
  \emph{Global atlas on cardiovascular disease prevention and control}.\hskip
  1em plus 0.5em minus 0.4em\relax World Health Organization, 2011.

\bibitem{mehra2007global}
R.~Mehra, ``Global public health problem of sudden cardiac death,''
  \emph{Journal of electrocardiology}, vol.~40, no.~6, pp. S118--S122, 2007.

\bibitem{yang2018automatic}
W.~Yang, Y.~Si, D.~Wang, and B.~Guo, ``Automatic recognition of arrhythmia
  based on principal component analysis network and linear support vector
  machine,'' \emph{Computers in biology and medicine}, vol. 101, pp. 22--32,
  2018.

\bibitem{raj2018sparse}
S.~Raj and K.~C. Ray, ``Sparse representation of ecg signals for automated
  recognition of cardiac arrhythmias,'' \emph{Expert systems with
  applications}, vol. 105, pp. 49--64, 2018.

\bibitem{li2019automated}
F.~Li, Y.~Xu, Z.~Chen, and Z.~Liu, ``Automated heartbeat classification using
  3-d inputs based on convolutional neural network with multi-fields of view,''
  \emph{IEEE Access}, vol.~7, pp. 76\,295--76\,304, 2019.

\bibitem{zhai2018automated}
X.~Zhai and C.~Tin, ``Automated ecg classification using dual heartbeat
  coupling based on convolutional neural network,'' \emph{IEEE Access}, vol.~6,
  pp. 27\,465--27\,472, 2018.

\bibitem{kiranyaz2015real}
S.~Kiranyaz, T.~Ince, and M.~Gabbouj, ``Real-time patient-specific ecg
  classification by 1-d convolutional neural networks,'' \emph{IEEE
  Transactions on Biomedical Engineering}, vol.~63, no.~3, pp. 664--675, 2015.

\bibitem{xu2020ecg}
X.~Xu and H.~Liu, ``Ecg heartbeat classification using convolutional neural
  networks,'' \emph{IEEE Access}, vol.~8, pp. 8614--8619, 2020.

\bibitem{sellami2019robust}
A.~Sellami and H.~Hwang, ``A robust deep convolutional neural network with
  batch-weighted loss for heartbeat classification,'' \emph{Expert Systems with
  Applications}, vol. 122, pp. 75--84, 2019.

\bibitem{zhang2017patient}
C.~Zhang, G.~Wang, J.~Zhao, P.~Gao, J.~Lin, and H.~Yang, ``Patient-specific ecg
  classification based on recurrent neural networks and clustering technique,''
  in \emph{2017 13th IASTED International Conference on Biomedical Engineering
  (BioMed)}.\hskip 1em plus 0.5em minus 0.4em\relax IEEE, 2017, pp. 63--67.

\bibitem{singh2018classification}
S.~Singh, S.~K. Pandey, U.~Pawar, and R.~R. Janghel, ``Classification of ecg
  arrhythmia using recurrent neural networks,'' \emph{Procedia computer
  science}, vol. 132, pp. 1290--1297, 2018.

\bibitem{rana2019ecg}
A.~Rana and K.~K. Kim, ``Ecg heartbeat classification using a single layer lstm
  model,'' in \emph{2019 International SoC Design Conference (ISOCC)}.\hskip
  1em plus 0.5em minus 0.4em\relax IEEE, 2019, pp. 267--268.

\bibitem{quionero2009dataset}
J.~Quionero-Candela, M.~Sugiyama, A.~Schwaighofer, and N.~Lawrence, ``Dataset
  shift in machine learning.(01 2009),'' \emph{Google Scholar Google Scholar
  Digital Library Digital Library}, 2009.

\bibitem{goldberger2017clinical}
A.~L. Goldberger, Z.~D. Goldberger, and A.~Shvilkin, \emph{Clinical
  electrocardiography: a simplified approach e-book}.\hskip 1em plus 0.5em
  minus 0.4em\relax Elsevier Health Sciences, 2017.

\bibitem{acharya2007advances}
U.~R. Acharya, S.~M. Krishnan, J.~A. Spaan, and J.~S. Suri, \emph{Advances in
  cardiac signal processing}.\hskip 1em plus 0.5em minus 0.4em\relax Springer,
  2007.

\bibitem{ye2022ecg}
Y.~Ye, T.~Luo, W.~Huang, Y.~Sun, and L.~Li, ``Ecg-based cross-subject mental
  stress detection via discriminative clustering enhanced adversarial domain
  adaptation,'' in \emph{2022 16th IEEE International Conference on Signal
  Processing (ICSP)}, vol.~1.\hskip 1em plus 0.5em minus 0.4em\relax IEEE,
  2022, pp. 495--499.

\bibitem{sagawa2019distributionally}
S.~Sagawa, P.~W. Koh, T.~B. Hashimoto, and P.~Liang, ``Distributionally robust
  neural networks for group shifts: On the importance of regularization for
  worst-case generalization,'' \emph{arXiv preprint arXiv:1911.08731}, 2019.

\bibitem{wang2021inter}
G.~Wang, M.~Chen, Z.~Ding, J.~Li, H.~Yang, and P.~Zhang, ``Inter-patient ecg
  arrhythmia heartbeat classification based on unsupervised domain
  adaptation,'' \emph{Neurocomputing}, vol. 454, pp. 339--349, 2021.

\bibitem{niu2020deep}
L.~Niu, C.~Chen, H.~Liu, S.~Zhou, and M.~Shu, ``A deep-learning approach to ecg
  classification based on adversarial domain adaptation,'' in
  \emph{Healthcare}, vol.~8, no.~4.\hskip 1em plus 0.5em minus 0.4em\relax
  MDPI, 2020, p. 437.

\bibitem{goldberger2000physiobank}
A.~L. Goldberger, L.~A. Amaral, L.~Glass, J.~M. Hausdorff, P.~C. Ivanov, R.~G.
  Mark, J.~E. Mietus, G.~B. Moody, C.-K. Peng, and H.~E. Stanley, ``Physiobank,
  physiotoolkit, and physionet: components of a new research resource for
  complex physiologic signals,'' \emph{circulation}, vol. 101, no.~23, pp.
  e215--e220, 2000.

\bibitem{sun2016deep}
B.~Sun and K.~Saenko, ``Deep coral: Correlation alignment for deep domain
  adaptation,'' in \emph{Computer Vision--ECCV 2016 Workshops: Amsterdam, The
  Netherlands, October 8-10 and 15-16, 2016, Proceedings, Part III 14}.\hskip
  1em plus 0.5em minus 0.4em\relax Springer, 2016, pp. 443--450.

\bibitem{huang2020self}
Z.~Huang, H.~Wang, E.~P. Xing, and D.~Huang, ``Self-challenging improves
  cross-domain generalization,'' in \emph{Computer Vision--ECCV 2020: 16th
  European Conference, Glasgow, UK, August 23--28, 2020, Proceedings, Part II
  16}.\hskip 1em plus 0.5em minus 0.4em\relax Springer, 2020, pp. 124--140.

\bibitem{sun2016return}
B.~Sun, J.~Feng, and K.~Saenko, ``Return of frustratingly easy domain
  adaptation,'' in \emph{Proceedings of the AAAI conference on artificial
  intelligence}, vol.~30, no.~1, 2016.

\bibitem{hasani2020classification}
H.~Hasani, A.~Bitarafan, and M.~S. Baghshah, ``Classification of 12-lead ecg
  signals with adversarial multi-source domain generalization,'' in \emph{2020
  Computing in Cardiology}.\hskip 1em plus 0.5em minus 0.4em\relax IEEE, 2020,
  pp. 1--4.

\bibitem{jin2022multi}
Y.~Jin, Z.~Li, Y.~Liu, J.~Liu, C.~Qin, L.~Zhao, and C.~Liu, ``Multi-class
  12-lead ecg automatic diagnosis based on a novel subdomain adaptive deep
  network,'' \emph{Science China Technological Sciences}, vol.~65, no.~11, pp.
  2617--2630, 2022.

\bibitem{ben2013robust}
A.~Ben-Tal, D.~Den~Hertog, A.~De~Waegenaere, B.~Melenberg, and G.~Rennen,
  ``Robust solutions of optimization problems affected by uncertain
  probabilities,'' \emph{Management Science}, vol.~59, no.~2, pp. 341--357,
  2013.

\bibitem{duchi2021statistics}
J.~C. Duchi, P.~W. Glynn, and H.~Namkoong, ``Statistics of robust optimization:
  A generalized empirical likelihood approach,'' \emph{Mathematics of
  Operations Research}, vol.~46, no.~3, pp. 946--969, 2021.

\bibitem{hearst1996reexamining}
M.~A. Hearst and J.~O. Pedersen, ``Reexamining the cluster hypothesis:
  Scatter/gather on retrieval results,'' in \emph{Proceedings of the 19th
  annual international ACM SIGIR conference on Research and development in
  information retrieval}, 1996, pp. 76--84.

\end{thebibliography}
\end{document}